\documentstyle[12pt]{article}
\newcommand{\nl  }{ \newline }
\newcommand{\np  }{ \newpage }
\newcommand{\R   }{ \mbox{${\rm R   }$} }
\newcommand{\I   }{ \mbox{${\rm I   }$} }
\newcommand{\III }{ \mbox{${\rm III }$} }
\newcommand{\V   }{ \mbox{${\rm V   }$} }
\newcommand{\VI  }{ \mbox{${\rm VI  }$} }
\newcommand{\VIo }{ \mbox{${\rm VI}_0$} }
\newcommand{\VIh }{ \mbox{${\rm VI}_h$} }
\newcommand{\II  }{ \mbox{${\rm II  }$} }
\newcommand{\IV  }{ \mbox{${\rm IV  }$} }
\newcommand{\VII }{ \mbox{${\rm VII }$} }
\newcommand{\VIIo}{ \mbox{${\rm VII}_0$} }
\newcommand{\VIIh}{ \mbox{${\rm VII}_h$} }
\newcommand{\VIII}{ \mbox{${\rm VIII}$} }
\newcommand{\IX  }{ \mbox{${\rm IX  }$} }
\newcommand{\KS  }{ \mbox{${\rm KS  }$} }
\newcommand{\SL  }{ \mbox{${\rm SL  }$} }
\newcommand{\X}{\mbox{${\rm X}$} }
\newcommand{\Y}{\mbox{${\rm Y}$} }

\author{Martin Rainer and Hans - J\"urgen Schmidt}
\title{Inhomogeneous cosmological models
with homogeneous inner hypersurface geometry}
\date{}
\begin{document}
\maketitle

\bigskip

\centerline{Universit\"at Potsdam, Institut f\"ur Mathematik}
\centerline{Projektgruppe  Kosmologie,
      D-14415 POTSDAM, PF 601553, Germany}
\smallskip
\centerline{Tel.: +49-331-962745, +49-331-9771347; Fax: +49-331-7499-203}
\smallskip
\centerline{e-mail: mrainer@aip.de, hjschmi@rz.uni-potsdam.de}
\bigskip

\begin{abstract}
Space--times which allow a slicing into homogeneous spatial
hypersurfaces generalize the usual Bianchi models. One knows
already that in these models the Bianchi type may change with
time. Here we show which of the changes  really appear. \par
To this end we characterize the topological space whose
points are the  $3$-dimensional oriented homogeneous
Riemannian manifolds; locally isometric manifolds are
considered as same.
\end{abstract}

AMS number: 53 B 20 Local Riemannian geometry

\bigskip

\section{INTRODUCTION}
In the review ''Physics in an inhomogeneous universe''
Krasi\'nski [1] defines a solution of the Einstein equation to
be cosmological if it can reproduce the
Friedmann--Lemaitre--Robertson--Walker
metric by taking limiting values of arbitrary constants or
functions. In the carefully collected bibliography [2] he
classified them according to geometrical properties of the
classes of solutions.
The main ingredient is the existence of isometries,
and it were Collins [3] and Krasi\'nski [4] who taught the relativists to
distinguish
between ''intrinsic''
 (internal) and external isometries (a distinction
known to mathematicians for long).
He showed that a cosmological model composed of homogeneous
 and isotropic spatial hypersurfaces need not be a Friedmann
 model.
The essential new feature is that the model can continuously
 change in time
between closed ($k=1$) and open ($k\le 0$) hypersurfaces
(which is impossible for the Friedmann models).
In [4a], the  Stephani universe was generalized by allowing
a changing value $k$ in the generalized Friedmann model.
Analogously, in [4b] the property  of spherical symmetry was
 generalized.

A likewise generalization of the spatially flat ($k=0$)
 Friedmann models
is the  Szekeres class of solutions [5, 6] of the Einstein
equation;
all elements of this class possess internally flat spatial
hypersurfaces.

In [7] the analogous question (which was considered by
Krasi\'nski for
homogenous isotropic hypersurfaces)   was posed and partially
answered
for models with homogeneous but not necessarily isotropic
hypersurfaces.

It is the aim of the present article to continue the research
done in [7]. In [7] one can read (we repeat it because that journal is not
available to everybody):
"To begin with, we consider the easily tractable case of a change to type I:
For each type M there one can find a manifold $V_4$ such that for $t\leq 0$
the section $V_3(t)$ is flat and for each $t>0$ it belongs to type M ...
Applying this fact twice it becomes obvious that by the help of a flat
intermediate all Bianchi types can be matched together.
However, if one does not want to use such a flat intermediate
the transitions of one Bianchi type to another one become
a non-trivial problem."

Let us repeat the question: A generalized homogenous
cosmological model
is a space--time possessing a foliation into homogenous
spatial
hypersurfaces. In which manner the corresponding Bianchi type
can change with time ?
This question can be formulated in several versions; we list five of them.
Other versions of the problem can be obtained by restriction to space--times
satisfying certain energy conditions; this is already mentioned in the
introduction of Ref. 6, but here we do not deal with such versions.

Let us consider continuous one-parameter families
$(\Sigma_\lambda,g_\lambda)$,
$\lambda\in [0,1]$,
where $\Sigma_\lambda$ be an orientable smooth three-manifold,
and $g_\lambda$ a homogeneous Riemannian metric on it.
Since here we consider the homogeneous
three-geometry only locally,
we can study it on some open standard manifold, $\Sigma=\Sigma_\lambda$
for all $\lambda$.
Globally however, $\Sigma_\lambda$ may change (see Discussion).

\medskip

Problem 1:
Fix a Bianchi type X and a Bianchi type Y. Then,
does there exist a one-parameter family of three-geometries
$(\Sigma, g_\lambda)$ (continuous in the parameter $\lambda$)
such that $(\Sigma, g_0)$ is of Bianchi type X and
$(\Sigma, g_1)$ is of Bianchi type Y?

\medskip

Problem 2:
Fix a three-geometry $(\Sigma, g_0)$
of Bianchi type X and a three-geometry $(\Sigma, g_1)$  of Bianchi type Y.
Then, does there exist a one-parameter family of three-geometries
$(\Sigma, g_\lambda)$ (continuous in the parameter $\lambda$)
reproducing $(\Sigma, g_0)$ and $(\Sigma, g_1)$
at $\lambda=0$ and $\lambda=1$, respectively?

Clearly both these problems can be answered
"Yes, for all possible X and Y"
by applying the cited phrase from [7] with
$g_\lambda$ being flat e.g. for $\frac{1}{3}\leq\lambda\leq\frac{2}{3}$.

\medskip

Problem 3:
Fix a Bianchi type X and a Bianchi type Y. Then,
does there exist a one-parameter family of three-geometries
$(\Sigma, g_\lambda)$ (continuous in the parameter $\lambda$)
such that $(\Sigma, g_0)$ is of Bianchi type X and
$(\Sigma, g_\lambda)$ is of Bianchi type Y for $0<\lambda\leq 1$ ?

\medskip

Problem 4:
Fix a three-geometry $(\Sigma, g_0)$
of Bianchi type X and a three-geometry $(\Sigma, g_1)$  of Bianchi type Y.
Then, does there exist a one-parameter family of three-geometries
$(\Sigma, g_\lambda)$ (continuous in the parameter $\lambda$)
and some critical parameter $\lambda_c$
such that $(\Sigma, g_\lambda)$ is of Bianchi type X for
$0\leq\lambda\leq\lambda_c$ and
$(\Sigma, g_\lambda)$ is of Bianchi type Y for
$\lambda_c<\lambda\leq 1$ and $g_0$, $g_1$ coincide with the prescribed
geometries ?

\medskip

Problem 5:
Fix a three-geometry $(\Sigma, g_0)$
of Bianchi type X and a three-geometry $(\Sigma, g_1)$  of Bianchi type Y.
Then, does there exist a one-parameter family of three-geometries
$(\Sigma, g_\lambda)$ (continuous in the parameter $\lambda$)
such that $(\Sigma, g_\lambda)$ is of Bianchi type X for
$\lambda = 0$ and
$(\Sigma, g_\lambda)$ is of Bianchi type Y for
$0<\lambda\leq 1$ and $g_0$, $g_1$ coincide with the prescribed geometries ?

It holds: The set of homogeneous
Riemannian three-geometries
of any fixed Bianchi type is arcwise
connected. So Problems 3 and 4 are equivalent.
Even more, it turns out that for a pair of Bianchi types $(\X,\Y)$
solving Problem 3 and 4 the corresponding sets $S_{\X}$ and $S_{\Y}$ of
homogeneous Riemannian three-geometries are related by
$S_{\X}\subset \partial S_{\Y}$,
where $\partial S_{\Y}$ denotes the boundary of $S_{\Y}$.
Hence Problem 5 is also equivalent to Problems 3 and 4.

This is the problem which is resolved in the following.
If the answer with is "Yes" for $(X,Y)$ with $\X\neq\Y$,
then we also say that there is a transition
$\Y\to \X$ corresponding to a directed graph (see Appendix B).

\bigskip

\section{PRELIMINARIES}
\setcounter{equation}{0}
In this section, we introduce the necessary differential
geometric and
group theoretic notions, and we shortly review the literature
to related topics.

The classical book by Wolf [8] does not only deal with the
Riemannian and
pseudo-Riemannian spaces of constant curvature, it covers also
a lot of
results on local and global differential geometry including
symmetric
and homogeneous manifolds.

\bigskip

\subsection{Isometry groups of Riemannian spaces}

\bigskip

Karlhede and MacCallum [9] deal with the equivalence problem
which consists of the problem to find a procedure for deciding whether two
given Riemannian spaces are isometric.

Here and in the following we use the notion ''isometry
group'' in the meaning ''the connected component of the unity of the full
isometry group''.

Szafron [10] poses the
``program to classify cosmological models using intrinsic
symmetries''.
He mentions the following results:
\par
{\bf Proposition 1}:
For Bianchi types II and III, the internal isometry group is
always
$4$-dimensional, for Bianchi types VIII and IX it is
sometimes,
and for Bianchi types IV and VI$_h$ $(h\neq 0,1)$ it is never
$4$-dimensional.
\par
{\bf Proposition 2}:
The conformal flatness of a $3$-dimensional Riemannian
manifold is equivalent
to the vanishing of the Cotton--York tensor. For homogenous
spaces this is
equivalent to the validity of $R_{i[j;k]} =0$
where $[\quad]$ denotes antisymmetrization.

Bona and Coll [11] go a similar way:
They look for the isometry groups of $3$-di\-men\-sio\-nal
Rie\-man\-nian
me\-trics, they
 understand it as part of the pro\-gram
''pro\-blem of equi\-va\-lence of me\-trics''.

Limits of space--times are considered in [12]
using a coordinate--free approach,
cf. also the references cited in [12].
In [7, 13 - 16] a topology in the space of Lie algebras was
considered
which shall be put in relation to the limit of space--times in
sct. 3.

A topology in the space of real Lie algebras can be defined as
follows:
Let $A$ be a set of $d$-dimensional Lie algebras and $V$ be
the
$d^3$-dimensional real vector space. We define a subset
$B \subset V$ with $x \in B$ iff $x$ represents a set
$C^i_{jk}$
of structure constants for an element of $A$.
Then $A$ is defined to be  closed
iff $B$ is closed in the Euclidean topology of $V$.
(For details see appendix A.)

Example: If $A$ has the commutative algebra as only element,
then $A$ is a closed set, because $B$ consists of the origin
of $V$ only.
On the other hand Segal [16] showed
\par
{\bf Proposition 3}:
Compact semisimple Lie algebras represent  isolated points in
the space
of Lie algebras,  because of the definiteness of their Cartan
metric.

Results of [14] (including the complete topology for the
$3$-dimensional case)
are generalized in [15] (also completing the $4$-dimensional
case).

There exist several similar constructions in the space of Lie
groups.
Co\-nat\-ser [17] con\-sidered the In\"on\"u--Wigner
contractions
of the low-dimensional real Lie algebras.
Levy-Nahas [18] and Saletan [19] define
contractions
and deformations of Lie groups. These concepts are a little more general ones.
It holds [18]
\par
{\bf Proposition 4}:
There exist exactly two Lie groups which can be contracted to
the
Poincar\'e group: The de Sitter group $SO(4,1)$ and the
anti--de Sitter group $SO(3,2)$.
Both of them are semisimple, they have different signatures of
the Cartan metric.

It is pointed out in [15] that all these deformations and
contractions
within the category of finite-dimensional Lie algebras appear
in the topology of the resp. space of Lie algebras,
but the latter has more general limits, some of which
correspond
 neither to In\"on\"u--Wigner nor to Saletan contractions.

In [20], the Heisenberg group is defined as odd-dimensional
simply connected $2$--step nilpotent ($ [g,g] \ne 0, \quad
[g,[g,g]] =0 $) real Lie group with $1$-dimensional center.
For every odd dimension, the Heisenberg group is uniquely
 defined by this condition. In $3$ dimensions, its algebra is
of Bianchi type II, in
dimension $d>3$ its algebra differs from the algebra
II$^{(d)}$ considered in [14, 15],
defined as direct product of II with the $(d-3)$-dimensional
Abelian algebra.
(II$^{(d)}$ is the atom defined in [14]. It has a
$(d-2)$-dimensional
center and is also considered as the generalized
$3$-dimensional Heisenberg algebra.)

Kaplan [21] also discusses groups of the Heisenberg type;
for 3 dimensions his definition coincides with the usual one,
for higher dimensions it differs both from [20] and [14] which
can be seen
if one compares the dimension of the centers.

The $3$-dimensional Lie algebras are classified according to
Bianchi.
Ref. [22] represents the original citation for the Bianchi
types,
the book [23] is more widely known and refers to [22].
However, Lie [24], [25]  classified them some years earlier;
nevertheless, we keep calling them Bianchi types.
Cf. also [26] for this classification.

Refs. [27 - 29] and many further ones deal with cosmological
models
with isometry subgroup belonging to one of the Bianchi types.

\bigskip

\subsection{Local homogeneity}

\bigskip

Let us now come to the condition for homogeneity of a
Riemannian manifold.
The theorem of Ambrose and Singer [30] is popularized in [31]
and gives
conditions under which a Riemannian manifold is locally
homogeneous. \\
Let $g$ be the metrical tensor, $\nabla _L$ the corresponding
Levi--Civita
connection and $R$ its curvature tensor. Let $T$ be the
torsion tensor and
$\nabla  =   \nabla _L \ - \ T$.
The Ambrose--Singer equations (AS--equations) are the
following:
$$\nabla g=0 \qquad   \nabla T=0 \qquad   \nabla R=0 $$
Let us sum up the main ingredients valid for all connected
Riemannian
mani\-folds $V_n$:

\bigskip

{\bf Proposition 5} ([30], [31]):
\par 1. If the  $V_n$ is complete, simply connected and
locally homogeneous,
then it is homogeneous.
\par 2. If the  $V_n$ is homogenous,
then there exists a solution of the AS--equations.
\par 3. If there exists a solution of the AS--equations,
then the $V_n$ is locally homogeneous.

\medskip

As special case one gets the following:
In the set of connected simply connected and complete
Riemannian manifolds
the AS--equations can be solved if and only if the manifold is
homogeneous.

(Remark:
In Ref. [11], Bona and Coll give (for $n=3$) also necessary and sufficient
conditions under which a Riemannian space is homogeneous.
However, their approach is quite different because they do not use
the auxiliary torsion as Ambrose and Singer did.)

A further interesting special case is $T=0$. The AS--equations
can be solved
with $T=0$ iff the Riemann tensor is covariantly constant.
Such spaces are called locally symmetric. It follows from the
AS--theorem,
that locally symmetric spaces are locally homogeneous.

In [32] the spatial curvatures of all the Bianchi types are
calculated.
Refs. [33, 34] deal with homogeneous Riemannian spaces.
Milnor [34] gives a review of left--invariant metrics on Lie
groups;
he writes:
''In the $3$-dimensional case, the theory is essentially
complete.''
To our knowledge, it has not been completed yet; more
precisely:
All particular parts of the puzzle exist,
but they have not been put together to a picture.
Even the present article will not finish this task.

It is mentioned in [34], that  the curvature depends
continuously on the
structure constants and on the metric. This gives already a
hint how the
puzzle can be handled.

\bigskip

\subsection{Special properties of isometry groups and eigenvalues of the Ricci
tensor}

\bigskip

A Lie group is {\it unimodular} if and only if the
left--invariant and the
right--invariant Haar measures coincide.
Equivalently one can say: a Lie group is unimodular iff the
structure
constants of the corresponding Lie algebra are trace--free,
i.e.
$C^i_{ji} =0$. \par
In the $3$-dimensional case, the $6$ unimodular groups are
Bianchi types I, II, VI$_0$, VII$_0$, VIII,  and IX,
sometimes also called type A-algebras.

\medskip

A metric is called {\it bi--invariant} if it is simultaneously
a left--invariant and a right--invariant one. It holds:
Only unimodular Lie groups carry bi--invariant metrics.
More specifically we have \par
{\bf Proposition 6} ([34]):
\par
A bi--invariant metric exists on a Lie group if and only if
the group is
the direct product between a compact and a commutative group.

\medskip

In the $3$-dimensional case, this takes place only for Bianchi
types I and IX.
For Bianchi type I, every metric is bi--invariant.
For Bianchi type IX the metric leading to a space of constant
curvature
is bi--invariant.

\medskip

Now we list some results dealing with the signs of the
eigenvalues of the
Ricci tensor. It holds \par
{\bf Proposition 7} (Theorem 2.2. of ref. 34]):
A connected Lie group admits a left invariant metric where all
eigenvalues
of the Ricci tensor are positive if and only if the group is
compact with
finite fundamental group.\\
For this case, the metric can be chosen to be bi--invariant,
such that, simultaneously, all eigenvalues of the Ricci tensor
coincide.\\
In the $3$-dimensional case, this takes place for Bianchi
 type IX only,
then, if all eigenvalues of the Ricci tensor coincide,
the space is of constant curvature.

\medskip

{\bf Proposition 8} ([34]):
If the Lie group is not unimodular then one of the eigenvalues
of the
Ricci tensor is negative.

\medskip

A Lie group is called {\it nilpotent} if for the corresponding
Lie algebra $g$
the sequence
$$g \supset [g,g] \supset [g,[g,g]] \supset \dots $$
terminates at zero. \\
In $3$ dimensions, Bianchi types I and II are nilpotent.
It holds \par
{\bf Proposition 9} ([34]):
Every left--invariant metric on a
non--commutative nilpotent Lie group has a positive and a
negative
eigenvalue of the Ricci tensor.

\medskip

A Lie group  is called {\it non--trivial} if there exist $3$
linearly
independent vectors $x$, $y$, $z$ of the corresponding Lie
algebra
such that $[x,y]=z$.
(A commutative Lie group is always trivial.)
In $3$ dimensions, Bianchi type V represents the only trivial
non--commutative Lie group. In analogy to proposition 9 we
have
\par
{\bf Proposition 10} ([34]):
Every non--trivial Lie group possesses a left--invariant
metric which
has a positive and a negative eigenvalue of the Ricci tensor.

In $3$ dimensions on gets furthermore: \par
{\bf Proposition 11}:
A left--invariant metric on a trivial Lie group has never a
positive
eigenvalue of the Ricci tensor.\\
Proof: Bianchi type I has vanishing curvature,
Bianchi type V leads always to a space of constant negative
curvature.

\medskip

{\bf Proposition 12} ([34]):
Every non--commutative Lie group carries a left--invariant
metric with
negative Ricci scalar.
\medskip

Partial results for the curvature invariants of Bianchi types
II and VIII
have been obtained independently in [35], a systematic
classification for the curvature
invariants of all Bianchi types can be found in [15].

\bigskip

\subsection{Homogeneous spaces and triviality}

\bigskip

In defining homogeneous manifolds we follow
 Ziller [37]. A Riemannian or pseudo-Riemannian manifold $M$
is
homogeneous if the isometry group
acts transitively
on it. It holds

{\bf Proposition 13}:
Each homogeneous manifold $M$ can be represented as $M=G/H$
 where $G$ is the isometry group of $M$ and $H$ its isotropy
subgroup.

Of course, sometimes a homogeneous manifold can be represented
as $M={\tilde G}/{\tilde H}$ where ${\tilde G}$ and ${\tilde
H}$ are not the isometry resp. isotropy groups.

Slansky [36] gives a review on
Lie groups. The exceptional Lie group $G_2$ is the only
$14$-dimensional
compact simple Lie group.
One gets the homogeneous sphere S$^6$ by factorizing
$G_2 / SU(3) = $S$^6$. This is a reduced quotient as compared to the
usual $SO(7)/SO(6)= $S$^6$,
where $ \dim SO(7)=21$.
In [31] this reads as follows:
The homogenous S$^6$ has two different homogeneous structures
with different Lie algebras.

Even more remarkable is also the following example:
The Heisenberg group (Bianchi type II) has two different
homogenous
structures, both with the {\em same} Lie algebra.
(The reason here is: Each left-invariant metric of the
Heisenberg group
possesses a $4$-dimensional isometry group, which is of type
$A_{4,10}$
in the classification of [15]. It admits two
different, though isomorphic, $3$-dimensional transitive
subgroups.)

Any $M=G/H$ is defined only with $H$ a closed subgroup
 of $G$. For a positive definite metric,
 moreover, $H$ is compact,
because it is a closed subgroup of the compact  rotation
group.
For this case one can do the following:
Let $g$ be the Lie algebra to $G$, $h$ to $H$, then there
exists a
subalgebra $p \subset g$ such that $g=h \oplus p$ and
$[h,p] \subset p$, so that $p$ can be identified with the
tangent space of $M$, cf. [37].

\medskip

Here we want to emphasize that this identification is not
automatically
possible for Lorentzian homogeneous metrics. In the following,
 we only deal with simply connected complete Riemannian
manifolds,
so we need not distinguish between homogeneous and locally
homogeneous spaces,
cf.   proposition 5.

\medskip

Further papers on related topics are [38] for space-time
homogeneous
manifolds, [39] is a continuation of ref. [11] for Lorentzian
metrics,
[40] discusses consequences of the fact that the Lorentz group
is not
compact (in contrast to the compactness of the  rotation
group).
[41] considers left--invariant Lorentz metrics on Lie groups.

A Lie algebra $g$ is {\em trivial} if for all $x$, $y \in g$
the
vector $[x,y]$ is a linear combination of $x$ and $y$.
(Of course, a Lie group is trivial iff the corresponding Lie
algebra
is trivial.)
Equivalently it holds:
$g$ is trivial and non--commutative if there exist a
commutative
subalgebra $h$ of codimension $1$ and a vector $b \in g
\backslash h$
such that for all $x \in h$ one has $[b,x]=x$.
A further equivalent definition is:
$g$ is trivial and non--commutative if there exists a basis
$e_1$, $e_2, \dots e_d$
such that for all $i,j = 2, \dots d$ one has
$[e_1,e_i]=e_i$ and $[e_i,e_j] = 0$. \\
Therefore, in each dimension exactly one non-Abelian trivial
algebra exists,
which is in fact identical to the unique non-Abelian {\em pure
vector type}
algebra of [14,15]. In 3 dimensions it is Bianchi type V.
Analogously to proposition 11 it holds

{\bf Proposition 14} ([41]):
Every left--inva\-ri\-ant me\-tric on a tri\-vial and
non--com\-mu\-ta\-tive
Lie group is of con\-stant ne\-ga\-tive cur\-va\-ture.\\
In $3$ dimensions this implies: Bianchi type V represents
always a $3$-space
of constant negative curvature, hence, it possesses a
$6$-dimensional
isometry group.

Concerning left--invariant Lorentz metrics one gets [41]:
For trivial Lie groups one gets only spaces of constant
curvature.
In $3$ dimensions it holds: Bianchi types II, VI$_0$ and
VII$_0$ admit
flat spaces, semi--simple algebras do {\it not} admit
flat spaces.

\bigskip

\section{HOMOGENEOUS  MANIFOLDS}
\setcounter{equation}{0}

In subsection 3.1 we deal with $2$-dimensional homogeneous
manifolds of
both possible signatures, in subsection 3.2
we consider the positive definite $3$-dimensional homogeneous
manifolds.

\medskip

\subsection{Two--dimensional homogeneous manifolds}
This subsection repeats only known results but the point of
view seems
to be new.
However, the main reason why we present it here is
to show the reader what style of reasoning is behind
subsection 3.2.
 We do not distinguish between manifolds whose only
difference
is an overall factor $(-1)$ in front of the metrical tensor.
So we have to consider only two possible signatures.

\subsubsection{Positive definite signature}
Let us start with the two $2$-dimensional Lie algebras.
Every left--invariant metric on the commutative Lie group is
flat.
This gives the flat Euclidean plane R$^2$ with isometry group
$E(2)$
belonging to Bianchi type VII$_0$. \par
The only non--commutative $2$-dimensional Lie algebra is
trivial
(cf. propositions 10, 11 and 14), and so every left--invariant
metric
gives the H$^2$, the $2$-surface of constant negative
curvature.
Its isometry group is $SO(2,1)$ and belongs to Bianchi type
VIII.  \par
What about the sphere S$^2$, the $2$-surface of constant
positive curvature?
Its isometry group is $SO(3)$ and it  belongs to Bianchi type
IX.
In the language of proposition 13 we can write
 S$^2=SO(3)/SO(2)$,
and the above consideration shows: The $3$-dimensional
isometry group
of S$^2$ does not possess a $2$-dimensional transitive
subgroup,
but the $3$-dimensional isometry group of H$^2$ has a
$2$-dimensional
transitive subgroup.

\bigskip

\subsubsection{Indefinite signature}
The commutative Lie group gives the flat $1+1$-dimensional
Minkowski space--time $M^2$. The isometry group is the
$2$-dimensional pseudo-Euclidean group of motions, $E(1,1)$,
with corresponding Bianchi type VI$_0$.

\par
The non--commutative Lie group gives
a space of constant non--vanishing curvature (sometimes called
$2$-dimensional de Sitter space--time $\bar S^2$), the
 iso\-metry group belongs to Bianchi type VIII. \par
Contrary to the case of  definite signature, a third type does not
exist.

\bigskip

\subsubsection{The algebras of the isometry groups}
Figure 1 represents a subdiagram of the diagram of refs. [7,
14], cf. also appendix A.
An arrow denotes convergence in the topology [14, 16], cf.
appendix B.

\bigskip

{\LARGE
$$\ \ \IX \ \longrightarrow \ \VIIo$$
$$\quad \  \nearrow$$
$$\ \VIII \ \longrightarrow \ \VIo$$
}

\medskip

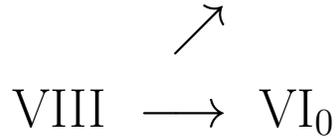
Fig. 1 First subdiagram for the Bianchi types

\medskip

One can see: The Lie algebra VI$_0$ can be deformed to type
VIII only,
cf. subsection 3.1.2. The Lie algebra
VII$_0$ can be deformed to types VIII and IX, cf. proposition
4 and
subsection 3.1.1.
(In accordance with proposition 3 the types VIII and IX
cannot be deformed or contracted into each other.)\\
So already at the Lie algebra level the difference between
3.1.1 and 3.1.2 becomes obvious.

\medskip

\subsubsection{The corresponding limits of space--times}
Now we come to the scope of this article: models with
intrinsically
homogeneous hypersurfaces. Here we consider the
$2+1$-dimensional case with $2$-dimensional hypersurfaces.
\np\noindent
{\LARGE
$$\ \ S^2 \ \longrightarrow \ R^2$$
$$\quad \  \nearrow$$
$$\ \ H^2 \qquad \qquad $$
$$\ \ \bar S^2 \ \longrightarrow \ M^2$$
}

\medskip

Fig. 2 Homogeneous $2$-surfaces of constant curvature; upper
part: definite signature; lower part: indefinite signature.

\medskip
Fig. 2 is to be read as follows:
There exists a sequence of spheres which converge locally to
the plane (by blowing them up), analogously there exists a
sequence of H$^2$'s which converges to the flat plane.
(But a direct continuous deformation
between S$^2$ and H$^2$ is not possible.) For indefinite
signature it holds analogously: There exists a sequence of
$\bar S^2$'s  which converge locally to the M$^2$.\\
It should be noticed that Fig. 2 gives the same topological
relation as the corresponding Lie group diagram  Fig. 1.
A $3$-dimensional space--time possessing a slicing into
$2$-dimensional homogeneous positive definite hypersurfaces
has  the following property:
A continuous change of the type of the slices is possible if
and only if the corresponding Lie algebras of the internal
isometry groups can be contracted into each other.
\par
If one allows complex transformations, the Lie groups \IX and
VIII on the one hand, and \VIIo and \VIo on the other hand,
become equivalent, cf. Fig. 1. Analogously,
S$^2$, H$^2$, and $\bar S^2$ on the one hand, and R$^2$ and
M$^2$ on the other hand, are isometric by a complex coordinate
transformation, cf. Fig.~2.
\par
The slice is locally uniquely determined by the value of its
curvature scalar and the signature of the metric. A
continuous change of the slice is possible iff the signature
is the same and the  corresponding inner curvature scalar
changes continuously.
\par
In the next subsection we want to clarify,
to which extent analogous  results are valid also in the more
interesting $3+1$-dimensional case.

\medskip

\subsection{Three--dimensional homogeneous manifolds}
We only consider the case of a positive definite metric.

\medskip

\subsubsection{The Kantowski--Sachs model}
Let us start with the Kantowski--Sachs model, whose spatial
hypersurfaces
are S$^2 \times \R$, the Cartesian product of the sphere and
the real line.
The isometry group is $4$-dimensional, and it does not possess
a $3$-dimensional transitive subgroup.
\par
The eigenvalues of the Ricci tensor are $(\lambda , \lambda,
0)$
where $\lambda $ is any positive real.
It is essential to note that this is the only case
(more exactly: a one--parameter set of cases, parametrized by $\lambda$)
for a
$3$-dimensional
homogeneous manifold in which no $3$-dimensional transitive
subgroup of the
isometry group exists.
A little less known is the following:
It is also the only homogeneous $3$-dimensional manifold where
the
Ricci tensor has eigenvalues  $(\lambda , \lambda, 0)$ with
$\lambda > 0$.
\par
Proof: Because of propositions 8 and 9 we have to check only
the unimodular
but not nilpotent Lie algebras, i.e. types
VI$_0$, VII$_0$, VIII and IX. The Ricci scalar equals
$2 \lambda > 0$, but only type IX allows a positive Ricci
scalar.
The rest is done by explicit calculus.

Besides this exceptional case one gets all other
homogeneous
3--spaces by considering the left--invariant metrics on
$3$-dimensional
Lie groups which are discussed in the next subsection.
Nevertheless it is useful to mention the following:
H$^2 \times \R$, the Cartesian product of the 2--surface of
constant
negative curvature and the real line possesses a
$4$-dimensional
isometry group; it has a transitive subgroup of Bianchi type
III
(which is the same as VI$_1$).
The Ricci tensor has eigenvalues  $(\lambda , \lambda, 0)$
with $\lambda < 0$.

\medskip

\subsubsection{The Bianchi models}
Let us consider now $3$-dimensional homogeneous spaces
different from the  Kantowski-Sachs spaces of the last section. These
so--called Bianchi models are
defined by left--invariant metrics on a
$3$-dimensional Lie group $G_3$. This is a simply transitive
isometry subgroup given by one of the Bianchi types.
{}From subsection 3.2.1. it follows that
subsections 3.2.1 and 3.2.2. together cover all
3--dimensional homogeneous Riemannian manifolds.

The different Bianchi types are as follows, cf. also appendix
A. The Lie algebra $A_3$ of $G_3$ is determined by the
commutators
\begin{equation}
[e_a,e_b]=C^c_{ab}e_c
\end{equation}
of its generators $e_a$, which may be represented
w.r.t. coordinates $\{x^\alpha\}$ as
left-invariant vector fields,
\begin{equation}
e_a=e^\alpha_a \frac{\partial}{\partial x^\alpha}.
\end{equation}
Their duals are the triad $1$-forms
\begin{equation}
e^a=e^a_\alpha dx^\alpha.
\end{equation}
The coordinate components
of the first fundamental form
$d^2s=\eta_{\alpha\beta}dx^\alpha{dx^\beta}$
may be expressed
by the anholonomic components w.r.t. the triad, i.e.
\begin{equation}
\eta_{\alpha\beta}=g_{ab}e^a_{\alpha}e^b_{\beta}.
\end{equation}
While $\eta_{\alpha\beta}$ depend in general on  $x^\alpha$,
we can choose the anholonomic representation without loss of
generality
such that $g_{ab}$ is constant and diagonal, i.e.
\begin{equation}
(g_{ab})=
\left [\begin {array}{ccc} e^{x}&0&0\\\noalign{\medskip}\
0&{e^{y}}&0\\\noalign{\medskip}0&0&{e^{z}}\end {array}\right
],
\end{equation}
which corresponds to a positively oriented orthogonal
triad frame of reference,
where the eigenvalues $e^x,e^y,e^z$ define the scales of
measurement in
the $3$ orthogonal directions.

In  calculations below we also use the parameters
\begin{equation}
t:=x-z;\quad u:=y-z;\quad w:=y-x.
\end{equation}

The  structure constants can be calculated from the triad
and $d^2s$ by
\begin{equation}
C_{ijk}=d^2s([e_i,e_j],e_k),\qquad C^k_{ij}=C_{ijr}g^{rk}.
\end{equation}

The  connection coefficients are given by
\begin{equation}
\Gamma^k_{ij}=\frac{1}{2}
g^{kr}(C_{ijr}+C_{jri}+C_{irj}),
\end{equation}
where
\begin{equation}
-\Gamma^k_{[ij]}=-\frac{1}{2}C^k_{ij}
=e^\mu_i e^\nu_j\partial_{[\mu} e^k_{\nu]}
\end{equation}
is the object of anholonomity.

The curvature operators to the triad basis are defined as
\begin{equation}
\Re_{ij}:=\nabla_{[e_i,e_j]}
     -(\nabla_{e_i}\nabla_{e_j}-\nabla_{e_j}\nabla_{e_i})
\end{equation}
and the Riemann tensor components are
\begin{equation}
R_{hijk}:=<e_h,\Re_{ij}e_k>.
\end{equation}
The Ricci tensor then is
\begin{equation}
R_{ij}:=R^k_{ikj}
=\Gamma^f_{ij}\Gamma^e_{fe}-\Gamma^f_{ie}\Gamma^e_{fj}
	 +\Gamma^e_{if}C^f_{ej} .
\end{equation}

The Ricci curvature scalar is $R:=R^i_{\ i}$,
the sum of the squared eigenvalues of the Ricci tensor is
\begin{equation}
N:=R^i_{\ j} R^j_{\ i}
\end{equation}
$S^i_{\ j}$ is the trace--free part  of the Ricci tensor, we
define
\begin{equation}
S:=S^i_{\ j} S^j_{\ k} S^k_{\ i}= R^i_{\ j} R^j_{\ k} R^k_{\
i} - R N + \frac{2}{9} R^3,
\end{equation}
and
\begin{equation}
Y:=\frac{1}{2}
(R_{ij;k}-R_{ik;j})g^{il}g^{jm}g^{kn}R_{lm;n}
\end{equation}
These are the four scalar invariants which characterize the
local homogeneous space.

This means that the bound "$p \le 3$'' of theorem 6 in ref.
 11 can be lowered to "$p \le 1$" if one restricts to the set of homogeneous
spaces. In other words: If one knows these four numbers, then the geometry of
the homogeneous space is uniquely determined (up to possible global
identifications).
 If the Bianchi type is given and is neither \VIII nor \IX, then the three
 eigenvalues of the Ricci tensor suffice to determine the local geometry
 completely. Both Bianchi types \VIII and \IX possess examples where this is
 not true.

\bigskip

There is a one--to--one correspondence between the ordered
tripel of numbers $(R, N, S)$ and the non-ordered tripel
 consisting of the three eigenvalues of the Ricci tensor.
 It depends on the situation which of these tripels is more
appropriate.

\bigskip

$N = 0$  appears for the flat space only.
For $N\neq 0$, the invariant $N$ is positive
and can be used for a homothetic rescaling  of
the metric,
\begin{equation}
{\hat g_{ij}}:=\sqrt{N} g_{ij},
\end{equation}
One gets for the Ricci tensor
$\hat R^i_{\ j} = R^i_{\ j} / \sqrt{N}$
and for the scalar invariants
\begin{equation}
{\hat N} = 1,
\quad
{\hat R} = R/{\sqrt{N}},
\quad
{\hat S} = S/N^{3/2},
\quad
{\hat Y} = Y/N^{3/2}.
\end{equation}

If two eigenvalues of the {Ricci} tensor are equal,
i.e., its matrix reads
\begin{equation}
Ric=
\left [
\begin {array}{ccc}        a&0&0
\\\noalign{\medskip}       0&a&0
\\\noalign{\medskip}       0&0&ad  \end {array}
\right ],
\end{equation}
we obtain
for $a{>\atop<}0$ respectively

\begin{equation}
\hat R =
\pm  \frac{d+2}{\sqrt{d^2+2}}, \qquad
 \hat S =
\pm \frac{2}{9} \ \frac{(d-1)^3}{(d^2+2)^{3/2}}
\end{equation}
For $d\in \R\cup\{\pm\infty\}$ the corresponding values
in the ${\hat R}$-${\hat S}$-plane lie on a double line $L_2$,
defined by the range $\vert\hat R\vert\leq\sqrt{3}$
and the algebraic equation
\begin{equation}
{162}{\hat S^2}=(3-\hat R^2)^3.
\end{equation}
cf. Fig. 3. All other algebraically possible points of the
${\hat R}$-${\hat S}$-plane lie inside the region surrounded
by the line $L_2$.

If one eigenvalue of the Ricci tensor equals $R$, i.e.,
if there exists a pair of  eigenvalues of the Ricci tensor
$(a,-a)$,
\begin{equation}
Ric=
\left [
\begin {array}{ccc}        a&0&0
\\\noalign{\medskip}       0&-a&0
\\\noalign{\medskip}       0&0&ad  \end {array}
\right ],
\end{equation}
we obtain
\begin{equation}
{\hat R}=
{ \frac{d}{\sqrt{d^2+2}} }, \qquad
{\hat S}=
 \frac{2}{9} \ \frac{d (d^2 - 3)}{(d^2+2)^{3/2}}.
\end{equation}
For $d\in \R\cup\{\pm\infty\}$ the corresponding values
in the ${\hat R}$-${\hat S}$-plane lie on a line $L_{+-}$,
defined by the range $\vert\hat R\vert\leq 1$
and the algebraic equation
\begin{equation}
{\hat S}=\frac{11}{9}\hat R^3-\hat R.
\end{equation}

\bigskip

\bigskip

Fig. 3 The algebraically possible values of $\hat R$ and
$\hat S$, explanation see text.

\bigskip

In the case that one eigenvalue of the Ricci tensor is zero,
\begin{equation}
Ric=
\left [
\begin {array}{ccc}        0&0&0
\\\noalign{\medskip}       0&a&0
\\\noalign{\medskip}       0&0&ad  \end {array}
\right ],
\end{equation}
we obtain for $a{>\atop<}0$ respectively
\begin{equation}
{\hat R}=
\pm  \frac{d+1}{\sqrt{d^2+1}}, \qquad
{\hat S}=
\pm \frac{1}{9}\frac{(d-2)(2d-1)(d+1)} {(d^2+1)^{3/2}}.
\end{equation}
For $d\in \R\cup\{\pm\infty\}$ the corresponding values
in the ${\hat R}$-${\hat S}$-plane lie on a line $L_{0}$, cf.
Fig. 3, defined by the range $\vert\hat R\vert\leq\sqrt{2}$
and the algebraic equation
\begin{equation}
{\hat S}=\frac{\hat R}{2}(1-\frac{5}{9}\hat R^2).
\end{equation}

\bigskip

Now we calculate the invariants to the homogeneous spaces
of the different Bianchi types from
their structure constants. Bianchi type I gives only the flat
space, but all other ones can be given in their normal form
according to eqs. (3.16, 17), and so be inserted into fig. 3.
Homogeneous spaces possessing a six--dimensional isometry
group are represented by the two tip points at
$\hat R = \pm \sqrt 3$. Proposition 6 implies that a subset
of
Bianchi type IX is at $\hat R = \sqrt 3$.
{}From proposition 14 one obtains that Bianchi type V is always at
$\hat R = - \sqrt 3$. \par
Homogeneous spaces which have a five--dimensional isometry
group do not exist. \par
Homogeneous spaces possessing a four--dimensional isometry
group are represented by points on the boundary line $L_2$
except the two tip points. From proposition 1 it follows that
Bianchi types II and III belong to this set.
(However, there are spaces located on $L_2$, which do not
possess a $4$-dimensional isometry group.)
 The eigenvalues of the Ricci tensor for Bianchi type II are
$(\lambda, - \lambda, - \lambda)$ with any
$\lambda > 0$, i.e.
\begin{equation}
{\hat R}_{\II}=
-{ \frac{1}{\sqrt{3}} }, \qquad
{\hat S}_{\II}=
\frac {16\,\sqrt{3}}{81}
\end{equation}
So, in the ${\hat R}$-${\hat S}$-plane Bianchi type \II
corresponds to the point
$(-{ \frac{1}{\sqrt{3}} },\frac {16\,\sqrt{3}}{81})$
where the line $L_2$  and the line $L_{+-}$
 meet in the quadrant $\hat R < 0 < \hat S$. \par

\medskip

Atoms [14] in the set of Lie algebras are those
non--commutative algebras which can be contracted to the
commutative algebra only.
For every dimension $d \ge 3$ there exist exactly two atoms,
for $d=3$ theses are Bianchi types II and V. In other words:
A 3--dimensional Lie algebra is an atom iff it is nilpotent or
trivial but not commutative. They represent the most simple
non--commutative Lie algebras. Together with Bianchi type IV
they represent the following figure, cf. appendix A.

\medskip
\np\noindent
\begin{center}
{\LARGE
II $\quad \longleftarrow \ $ IV

$\downarrow \qquad \qquad \downarrow$

I $\quad \longleftarrow \quad$ V
}
\end{center}
\medskip

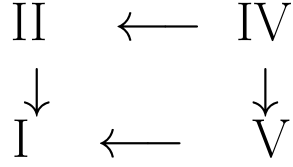
Fig. 4 Second subdiagram for the Bianchi types

\medskip

For Bianchi type IV we get
\begin{equation}
Ric_{\IV}=
\left [\begin {array}{ccc}
-{\frac {4\,{e^{w}}-1}
{2\,{e^{z+w}}}}
&
-\frac{1}{e^{z}}
&0\\\noalign{\medskip}
-{\frac {1}{{e^{z+w}}}}
&
-{\frac {4\,{e^{w}}+1}
{2\,{e^{z+w}}}}
&0\\\noalign{\medskip}0&0&
-{\frac {4\,{e^{w}}+1}
{2\,{e^{z+w}}}}
\end {array}\right ]
\end{equation}

\begin{equation}
N_{\IV}=
{\frac {48{e^{2w}}+16{e^{w}}+3}
{ 4\,{e^{2(z+w)}} }}
\end{equation}

\begin{equation}
{\hat R}_{\IV}=
-{\frac {12\,{e^{w}}+1}{\sqrt {48{e^{2w}}+16{e^{w}}+3}}}
\end{equation}

\begin{equation}
{\hat S}_{\IV}=
{\frac {8\left (9{e^{w}}+2\right )}
{9\,\left (48{e^{2w}}+16{e^{w}}+3\right )^{3/2}}}
\end{equation}

Thus, in the ${\hat R}$-${\hat S}$-plane the {Bianchi} type
\IV
corresponds to an open line (parametrized by $w$) connecting
the edge point $(-\sqrt{3},0)$ Bianchi type V (approached by
$w\to\infty$) and the {Bianchi} \II point (approached by
$w\to-\infty$). \par
If one removes the parameter $w$ one gets
$$\hat S = \frac{2}{3} \ (3-\hat R^2)^2 \
 \frac{5 - \frac{2}{3} \hat R^2 - \hat R \sqrt{18-5\hat R^2} }
{[\sqrt{18-5\hat R^2}- \hat R   ]^3}$$
a formula which is valid for $-\sqrt 3 < \hat R < - 1/\sqrt
3$.

\medskip

{}From subsection 3.2.1 we get for the Kantowski--Sachs
models always
$$\hat R = \sqrt 2 , \qquad \hat S = - \frac{\sqrt 2}{18}$$
and  the mirror point III is part of the
 Bianchi type III manifolds:
$$  \hat R = - \sqrt 2 , \qquad \hat S =  \frac{\sqrt 2}{18}$$
(Remark: It follows from proposition 10 that Bianchi type
III has also other manifolds.) These are the
points where the lines $L_2$  and $L_{0}$ intersect
 in the quadrants $\hat R \  \hat S < 0$. \par
In the ${\hat R}$-${\hat S}$-plane, to each of the {Bianchi}
types \VIh with $h\geq 0$ corresponds one line starting
from the point of II (for $w \longrightarrow \infty$, $w$
from eq. (3.6)).

For $h=1$ this curve is a segment of $L_2$, for $h\ne 0, \ 1$ it has no
common point with $L_2$. This behaviour follows already from proposition 1.
\par
For $h\longrightarrow\infty$ the \VIh--curve converges to
the IV--curve. The \VIo--curve is part of $L_{+-}$. \par
The \VIh ($0<h<1$)--curves cover that part of the areas A2 and A10 which lie
above the dotted line.
\par The \VIh ($h>1$)--curves cover all points above
the IV--curve which are either in area A2 and below $L_2$ or in A1 and below
or at the dotted line.

 A  $1$-parameter subset of \VIh (defined by $w=0$) corresponds to a curve
 (which is dotted in fig. 5)
 starting from the
point $(-1, -\frac{2}{9})$ (which is the point where the three lines $L_0$,
$L_{+-}$ and $L_2$ intersect at $\hat R < 0$, i.e. the other end of the
\VIo line, shortly  called
''point of \VIo''), passing the
point of \III for $h=1$ and approaching the point of \V for
$h\to\infty$. One gets
$$\hat R=-\sqrt{3-\frac{2}{h^2+1}}, \qquad
\hat S = \frac{18h^2-2}{9[(h^2+1)(3h^2+1)]^{3/2}}$$
Example: $h=\frac{1}{3}$ gives $\hat S = 0$ and $\hat R = -
\sqrt{6/5}$. In the diagram fig. 5 this line can be inserted
by eliminating the parameter $h$ to give
$$\hat S = - (3-\hat R^2)^2 \ \frac{5 \hat R^2 - 6}{18 \hat
R^3} $$
which makes sense for $-\sqrt 3 < \hat R < -1$.

\medskip

For {Bianchi} type \VIIh,
the expressions for the scalar invariants
$N$, $\hat R$, and $\hat S$
are symmetric under the parameter reflection
$w\leftrightarrow-w$.
For $w=0$, $h>0$ we obtain $(\hat R_{\VII},\hat
S_{\VII})=(-\sqrt{3},0)$,
which is the same as $(\hat R_{\V},\hat S_{\V})$. This
reflects the fact that the 6--dimensional isometry group of
Bianchi type V possesses 3--dimensional transitive subgroups
belonging to Bianchi type VII$_h$ for every value $h>0$. \par
In the ${\hat R}$-${\hat S}$-plane,
to each of the {Bianchi} types \VIIh
with fixed $h>0$
there corresponds an open line (parametrized by $w$)
between the points of \V (for $w=0$) and
\II (for $\vert w\vert\to\infty$).
These lines approach  the line of {Bianchi}
type \IV for $h\to\infty$ with the same endpoints. For $h\to
0$ they approach
the open piece of the line $L_{+-}$ which lies between
\II and the origin $(0,0)$. This line segment is the locus of
the {Bianchi} type \VIIo.
The lines of the types \VIIh, $h>0$, lie in the interior of a
region bounded by the lines of \VIIo, \IV and the axis $\hat
S=0$. The explicit expressions for the eigenvalues of the
Ricci tensor for types \VIh and \VIIh are complicated
 (see e.g. [15] or [32]), so we
give them for \VIIo only:
$$Ric \ \VIIo \sim (1-e^{2w},\ e^{2w}-1, \ -(1-e^w)^2)$$
One peculiarity must be mentioned: For positive $w$ the
corresponding points in the $\hat R$-$\hat S$-diagram converge
to the origin as $w$ tends to zero, but for $w=0$ the
resulting space is flat and cannot be represented in that
diagram. (The latter is a consequence of the
fact that the 6--dimensional isometry group
 of the flat space possesses a 3--dimensional
transitive subgroup of Bianchi type \VIIo.) II, \VIo and
\VIIo together fill the $\hat R < 0$-part of the
curve $L_{+-}$.
\par
The $2$-parameter expressions for the scalar invariants
$N$, $\hat R$, and $\hat S$ for {Bianchi} type \VIII and \IX
are symmetric under $u\leftrightarrow t$.
$$Ric \ \VIII \sim (1+2e^{t+u}-e^{2u}-e^{2t},
\ e^{2t}-1-e^{2u}-2e^u, \ e^{2u}-1-e^{2t}-2e^t)
$$
$$Ric \ \IX \sim (1+2e^{t+u}-e^{2u}-e^{2t},
\ e^{2t}-1-e^{2u}+2e^u, \ e^{2u}-1-e^{2t}+2e^t) $$
In the ${\hat R}$-${\hat S}$-plane, the
boundary of the regions of
{Bianchi} types \VIII
and \IX is given by the lines
$L_2$, $L_{0}$ and that of \VIIo, which is the
common boundary of \VIII and \IX. This boundary
can be reached from each region as an asymptotic limit.
{Bianchi} type  \VIII lies in the region to the left w.r.t.
\VIIo, while {Bianchi} \IX in the region to the right  w.r.t.
\VIIo.

The points of \II and $(0,0)$, the endpoints of the line
\VIIo, are corner points to both \VIII and \IX. They can be
reached
from each of them as asymptotic limit points.
The point \III is a corner point
of the region \VIII. There, one endpoint of $L_0$ hits on
$L_2$.

The point $(\sqrt{2},-\frac{1}{9\sqrt{2}})$ is a tip
of the region \IX which does not belong to \IX itself but can
be reached
as a limit of from points in \IX, e.g. for $(t,u)\to
(\infty,\infty)$.
This point corresponds to the {Kantowski-Sachs} spaces
(KS in Fig. 5 below), cf. subsection 3.2.1.

In the ${\hat R}$-${\hat S}$-plane
the region of \IX is connected.
In its point $(1,\frac{2}{9})$ (which, by the way, is the
mirror point to the \VIo point)
the lines $L_2$ and $L_0$ intersect.
If this point in the ${\hat R}$-${\hat S}$-plane were
missing, \IX would be disconnected. Note that this
point in the ${\hat R}$-${\hat S}$-plane  actually corresponds
to a $1$-dimensional line in the connected $2$-dimensional
space of homogeneous spaces modulo absolute scale.
It is then a special effect of the projection to the
${\hat R}$-${\hat S}$-plane that that $1$-dimensional
line degenerates to a point in this plane. Here the fourth
curvature invariant $Y$ eq. (3.15) is necessary.

\medskip

According to  [11],
if a maximal $4$-dimensional isometry group exists,
then it is either $\IX\oplus\R$, $\VIII\oplus\R$
or $A_{4,10}$.
This nicely agrees with possible transitions
$\IX\oplus\R\to A_{4,10}$ and $\VIII\oplus\R\to A_{4,10}$.

\medskip

Fig. 5 The homogeneous 3--spaces, a) with description
b) only the essential lines and the 16 areas \par

We conclude this section with Fig. 5
depicting the locus of local homogeneous 3--spaces
with positive definite metric in the  $\hat R$-$\hat S$-plane
defined by fig. 3.
The regions covered by type \VIII\ and \IX are horizontally
hatched, the region covered by lines \VIIh
for $h>0$ is vertically hatched.
The outer boundary line $L_2$ (double eigenvalue), the
 inner boundary
line $L_0$ (some zero eigenvalue) and the line $L_{+-}$ (pair
of eigenvalues
differing only in their sign) are drawn as full lines.
The dotted line is part of the boundary of the region covered by \VIh, $h>0$,
and \IV is a full
line.

\section{ORIENTATION}

If we prescribe an orientation, then both a Lie algebra and a
homogeneous manifold get additional structure. For
the 3--dimensional Lie algebras one can say the
following: Bianchi
types I, V, and \VIh (arbitrary $h$) are self--dual, i.e.,
there exists an orientation--reversing Lie algebra
isomorphism. All other Bianchi types lead to pairs of dual
algebras if the orientation is prescribed, cf. e.g. [14]. \par
 The analogous question for the corresponding manifolds is not
completely answered yet. But for cosmological models it should
be required that the spatial orientation keeps always the
same. To go further, one should find
 out, which of the
homogeneous 3--manifolds possess an orientation--reversing
isometry; we call such manifolds also self--dual. \par
 It holds: If the space is locally symmetric (cf. the  text
after  proposition 5) or if a hypersurface--orthogonal Killing
vector field exists, then such an isometry exists.
Therefore, all manifolds of Bianchi types I, II, III, V and
the KS--spaces, and some of the Bianchi IX manifolds
 are self--dual. \par
A further partial answer is as follows: If all eigenvectors of
the Ricci tensor are different and the underlying Lie algebra
of isometries is not self--dual, then also the manifold is not
self--dual. Therefore, all Bianchi type IV, some of Bianchi
type VII, some of VIII and some of type IX manifolds fail to
be self--dual.

\medskip

\section{DISCUSSION}
The answer to Problem 3 (see the end of the Introduction)
is "Yes" if and only if $(\X,\Y)$ is one of the following pairs:
$(\X,\Y)$ for $\X=\Y$,
$(\I,\Y)$ for arbitrary \Y,
$(\II,\Y)$ for $Y\not\in\{\I,\V\}$,
$(\V,\IV)$,
$(\VI_0,\VIII)$,
$(\VII_0,\Y)$ for $Y\in\{\VIII,\IX\}$.

Note that $\lim_{h\to\infty}\VI_h=\IV=\lim_{h\to\infty}\VII_h$,
hence Problems 1 and 2 have solutions by a continuous change of the parameter
$h$ in every subset of \{ \IV, \VIh, \VIIh \} also without going through the
flat intermediate.

If we generalize Problems 3, 4, 5 for Bianchi types
to the analogous Problems 3', 4', 5' for Bianchi and Kantowski-Sachs types
then 3', 4', 5' are again equivalent.
This problem has additional solutions for following pairs:
$(\KS,\IX)$,
$(\X,\KS)$ for $\X\in \{\I,\KS\}$.

Generalized homogeneous cosmological models enjoy a renewed
interest, see e.g. [1, 2, 42].

The review [43] entitled "Cosmic Topology" contains many valuable facts
on related questions. E.g. Thurston's homogeneous
three-geometries (appearing as universal
covering spaces of some compact homogeneous three-geometries
as quotients w.r.t. some discrete isometry subgroup acting simply transitive
on its orbits)
correspond to some characteristic Bianchi or Kantowski-Sachs types each.
(The reverse is not true.)
According to [43], $R^3$, $S^3$, $H^3$, $S^2\times R$, $H^2\times R$,
$\tilde{\SL}(2,R)$
correspond to type I, IX, V, KS, III, VIII, respectively.
We add: the remaining Thurston types,
$Nil$ and $Sol$, correspond to Bianchi types II and $\VI_0$ respectively.
So it is obvious, as mentioned in the Introduction, that  certain
transitions of the local geometry to another Bianchi or  Kantowski-Sachs type
determine changes of the global topology of the homogeneous
three-manifolds. Hence, although we concentrated here on a
classification problem of transitions in local geometry,
it also helps to classify related transitions of the global geometry.

In the present paper we
concentrated on the
question, how the Bianchi type can change with time in those
models, and how this is related to the topology in the space of
3--dimensional Lie algebras.
One should mention that the definition for the generalized
homogeneous cosmological models differs in different papers.
The main difference is the following:
''There exists a system of reference such that the spatial
slices $t=$ const. have  homogeneous inner geometry.'' versus
''There exists a {\it synchronized} system of reference such
that the spatial slices $t=$ const. have  homogeneous inner
geometry.'' (compare also Refs. [3, 4]). For our purpose this distinction is
not essential,
but if one solves the Einstein equation for models of this
kind the distinction  becomes essential.

A systematic consideration of homogeneous models is also
useful for a canonical quantization. In [32] the Wheeler--de
Witt  equation was considered for the
Bianchi models with $2$ or $3$ minisuperspace dimensions.
\par
What has to be done yet ? The analogous question should be
considered for Lorentz signature spaces. The set of manifolds
is no more known to be
a Hausdorff space, the calibration to $\hat N=1$ is
no longer possible in general, and more subcases have to be
distinguished. Also for positive definite signature, some
questions remain open yet, e.g.: ''If the four invariants $R$,
$N$, $S$, and $Y$ (eqs. (3.13-15)) of a sequence of
homogeneous 3--spaces converge, then the corresponding
manifolds also converge.'' This statement is stronger than
''Equality of these 4 invariants implies isometry.''
\par
The main result of the present paper are figures 3 and 5. The
 4 lines $L_0$, $L_{+-}$, axes $\hat R = 0$, and $\hat S = 0$ divide the
 surface surrounded by
$L_2$ of fig. 3 into $2^4=16$ areas. We denote the upper 8 areas
from left to right by A1, $\dots $, A8 and the lower 8 areas
from left to right by A9, $\dots $, A16. (So, area A$m$ and
area A$(17-m)$ are centrally mirrored to each other.) We have
shown at the beginning of sct. 3.2.2. that all these 16 areas
represent algebraically  possible values for $\hat R$,
$\hat S$. \par
 Now we can add: The interior of the following 7
areas does not represent a homogeneous space: A6, A7, A9, A12,
A13, A14, A15. Exactly four points of the boundaries of
 these areas represent
 homogeneous manifolds: KS, Bianchi type V, the \VIo point
and that type IX space with eigenvalues $(0, \ 0, \ 1)$. No point lying above
the dotted line in region A1 represents a homogeneous space. This
enumeration is complete: All other points in the diagram do
really represent a homogeneous space.
 (Up to now, one can find in the literature only examples
like: ''There does not exist a homogeneous $V_3$ whose
eigenvalues of the Ricci tensor
are $(\lambda, - \lambda, 0)$ with $\lambda \ne 0$.'' This
corresponds to the value $\hat R = \hat S = 0$.)  The
interior of the 4 areas A4, A5, A8, A16 is fully covered by
Bianchi type IX spaces. Only the 5 areas A1, A2, A3, A9, A10
possess a more detailed structure which becomes clear from
sct. 3. \par
Section 3.1. figs. 1 and 2 can be summarized as follows: If a
sequence of 2-dimensional homogeneous manifolds converges,
then the corresponding Lie algebras of
 the (always 3--dimensional!) isometry
groups also converge in the Segal
topology. Vice versa: If the Lie algebras converge in the
Segal topology, then one can find examples of correspondingly
converging manifolds.  \par
Section 3.2. figs. 3 - 5 does not possess such a simple
summary. The reason is that the dimension of the isometry
groups for 3--dimensional homogeneous manifolds varies between
3, 4 and 6. But for some subsets one can state analogous
results, let us pick up fig. 4, presenting Bianchi types I,
II, V, and IV. It holds within this set: If a sequence of
manifolds converges, then the corresponding Lie algebras
converge in the Segal topology, and analogous vice versa as
above. One can illustrate this situation in a plane where the
origin represents type I, the positive $x$--axis type V, the
positive $y$-axis type II, and the region $x>0, \ y>0$
 represents all type IV spaces. In other words: If a sequence
of manifolds of type IV converges to a homogeneous manifold of
another type, then it is of Bianchi type I, II, or V. Every
manifold of type I, II, and V appears as limit of such a sequence.
\medskip
\nl
\nl\noindent
{\Large
{\bf APPENDIX A. THE TOPO\-LO\-GI\-CAL \\* SPACE OF LIE
AL\-GE\-BRAS}}\\
\setcounter{equation}{0}
\nl
The bracket $[\ ,\ ]$ defines a Lie algebra of dimension $n$,
iff the structure constants
satisfy the $n\{{n\choose 2}+{n\choose 1}\}$ antisymmetry
conditions $C^k_{[ij]}=0$,
and the $n\cdot{n\choose 3}$ quadratic compatibility
constraints $C^l_{[ij}C^m_{k]l}=0$,
 called  Jacobi identity.

The space of all sets $\{C^k_{ij}\}$ satisfying the
antisymmetry condition and the Jacobi identity
can be considered as a subvariety  $W^n \subset \R^{n^3}$ of
dimension
$$
\dim W^n \leq n^3 - \frac{n^2(n+1)}{2}  =\frac{n^2(n-1)}{2}.
$$
For $n=3$ the structure constants can be written as
$$
C^k_{ij}=\epsilon _{ijl}(n^{lk}+\epsilon ^{lkm}a_m),
$$
where $n^{ij}$ is symmetric and
$\epsilon _{ijk}=\epsilon ^{ijk}$ totally
antisymmetric with $\epsilon _{123}=1$.
The Jacobi identity
is equivalent to $n^{lm}a_m=0$. Without loss of generality one gets
$a_m = (h, \ 0, \ 0)$
and $n^{lk}=diag(n_1, \ n_2, \ n_3)$. The
quadruples $(h, \ n_1, \ n_2, \ n_3)$ for the Bianchi types
are ($h \ge 0$ but arbitrary):
I$(0,0,0,0)$; II$(0,1,0,0)$; III$=$VI$_1$; \
IV$(1,0,0,1)$; \ V$(1,0,0,0)$; \ VI$_h$$(h,0,1,-1)$; \\*
VII$_h$$(h,0,1,1)$; VIII$(0,1,1,-1)$ and IX$(0,1,1,1)$.

Throughout the following, we will need the {\em separation
axioms} from topology. \\*
{\bf Axiom $T_0$}: For each pair of different points there is
an open set
containing only one of both.\\*
{\bf Axiom $T_1$}: Each one--point set is closed. \\*
{\bf Axiom $T_2$} ({\sc Hausdorff}):
Each pair of different points has a pair of disjoint open
neighbourhoods. \par
It holds: $T_2$ implies $T_1$, $T_1$ implies $T_0$.

$GL(n)$ basis transformations  induce
$GL(n)$ tensor transformations between equivalent structure
constants.
$$
C^k_{ij} \sim (A^{-1})^k_h\ C^h_{fg}\ A^f_i\ A^g_j \ \ \forall
A \in GL(n),
$$
where $\sim$ denotes the equivalence relation.
This induces the space $K^n=W^n/GL(n)$
of equivalence classes  w.r.t. the
nonlinear action of $GL(n)$ on $W^n$.

The ${GL}(n)$ action on $W^n$ is not free in general. It
holds:
$$
\dim W^n> \dim K^n \geq \dim W^n - n^2.
$$
The first inequality  is a strict one,
because the multiples of the unit matrix in $GL(n)$ give
rise to equivalent points of $K^n$.

Let $\phi: W^n\to K^n$ be the canonical map for the
equivalence relation $\sim$ defined by the action of $GL(n)$
in $W^n$. The natural topology $\kappa^n$ of $K^n$ is given as
the quotient topology of the induced topology of $W^n \subset
\R^{n^3}$ w.r.t. $GL(n)$ equivalence. This means: A sequence
of Lie algebras converges to a limit algebra if there exists a
basis for each algebra such that the corresponding structure
constants converge to each other as real numbers.
For $n\ge 2$, the topology  $\kappa^n$ is a $T_0$ but
 not a $T_1$--space. \par
Figure 6 shows the topological relation between the
Bianchi types, the relation between the graph and
the topology is outlined in appendix B. \\
\nl
\medskip
{\Large {\bf
APPENDIX B.
DI\-REC\-TED GRAPHS \\*
AND FI\-NITE TO\-PO\-LO\-GI\-CAL SPACES}}\\*
\nl
This appendix shall help reading the diagrams.
First we give the intuitive idea and second we outline the
mathematical apparatus behind it.

First, look at Fig. 4 and imagine that ''I'' is represented by the
origin of a Cartesian coordinate system.
''V'' shall denote all points of the positive $x$--axis,
''II'' the positive $y$--axis, and ''IV'' is the region of points
with $x>0$ and $y>0$. Then it is clear what the arrows mean:
the possible convergence of respective representatives.
There is no extra arrow directly from ''IV'' to ''I'' (seen
e.g. by representatives  at the line $x=y$), because the
transitivity property is know to be satisfied, and we want to have a
minimal number of arrows.

\bigskip

Second, let  $X$  be a finite nonvoid set.  A directed graph
 in $X$ is a subset
$\Gamma \,  \subset \,  X \,  \times \,  X$;  $(x,y)
\in
\Gamma$ means: there is a directed edge from vertex $x$ to
vertex
$y$.  The directed graph $\Gamma$ is called
transitively  closed,
if $(x,y) \in \Gamma$ and $(y,z) \in \Gamma$ imply
$(x,z) \in \Gamma$ and $(x,x) \in \Gamma$ for all $x \in X$.
 The  transitive  closure $Cl \,  \Gamma$ of the  directed
graph
$\Gamma$  is  the  smallest transitively  closed  directed
graph
containing $\Gamma$. Each  directed  graph   $\Gamma$  defines  a  topology
in $X$  as
follows:  $A \,  \subset \, X$ is closed iff $x \in A$ and
$(x,y)
\in \Gamma$ imply $y \in A$.

P r o o f: $\emptyset$ and $X$ are closed by this definition.
$X$
is   finite,   so  it  suffices  to  show  that  both  union
and
intersection of two closed subsets are closed. $\Box$

It holds:  Two directed graphs $\Gamma,  \, \tilde \Gamma$
define
the  same topology iff  $Cl \,  \Gamma  \,  = \,  Cl  \,
\tilde
\Gamma$.  The  topology  defined by $\Gamma$ is  finer  than
the
topology defined by $ \tilde \Gamma$  iff
 $Cl \,  \Gamma  \,  \subset \,  Cl  \,  \tilde \Gamma$.\\*
P r o o f:  For a directed graph, we define the
closure-operation
$cl_{\Gamma}$ as follows: For $A \subset X$,
$\quad cl_{\Gamma}(A)  \quad := \quad A \cup \{y\vert x \in A, (x,y)
\in \Gamma \}$ \\*
$cl_{\Gamma}$ is
idempotent (i.e., $cl_{\Gamma} \sp 2 \equiv
 cl_{\Gamma} \circ  cl_{\Gamma} = cl_{\Gamma}$) iff
$\Gamma  \,  \cup  \,  \{(x,x)\vert x \in X \}$  is
 transitively
closed.  Let  card $X = m$,  $A \subset X$ and $cl \,  A$ be
the
closure  of  $A$  in the topology defined by  $\Gamma$,  then
it
holds $cl \, A = cl_{\Gamma} \sp m A$.  $\Box$\\*
Examples:  The discrete topology in $X$ is defined from the
empty
graph  $\Gamma  =  \emptyset$.  The trivial topology  in  $X$
is
defined  from the complete graph $ \tilde \Gamma = X  \times
X$.
The inverse graph to $\Gamma$ is defined by
$$\Gamma \sp{-1} \, := \, \{(y,x)\vert (x,y) \in  \Gamma \} $$

It  holds:  The topologies defined by $\Gamma$ and
$\Gamma \sp{-
1}$ are inverse to each other,  i.e.,  $A \subset X$ is closed
in
the  topology defined by $\Gamma $ iff it is open in the
topology
defined by  $\Gamma \sp{-1}$.

Definition: A closed path in  $\Gamma$ is a sequence of edges
$$(x_1, x_2), \, (x_2,x_3), \dots , \, (x_n, x_1) \in \Gamma$$
connecting at least two vertices (i.e., card $\{x_1, \dots, \,
x_n\} \ge 2$). It holds:  The topology defined by $\Gamma$ is
$T_0$ iff
$\Gamma$ does not contain a closed path.\\
Let  us  now go the other direction:  Let a topology  in  $X$
be
given,  $cl$ denotes the closure with respect to it.  We
define a
directed graph $\Gamma$ by $(x,y) \in \Gamma$ iff
$y  \in  \,  cl\,  \{x\}$.  It  holds:  This  graph  $\Gamma$
is
transitively  closed,  and the topology defined from it
coincides
with the initial topology.

Conclusion:  There  is  a one-to-one correspondence  between
all
topologies  and  all transitively closed directed graphs  in
the
finite set $X$. \\
For  visualizing  a  finite  topological  space,   however,
the
transitively closed graphs are not best suited, one should
prefer
a graph with less edges. To find the best suited graph we
define

Definition: A directed graph $\Gamma$ is called minimal if
$\tilde \Gamma \subset \Gamma$ and $Cl \, \tilde \Gamma
\, = \,  Cl \, \Gamma$ imply $\tilde \Gamma \, = \, \Gamma$.

Each directed graph $ \Gamma $ contains a minimal subgraph
$\tilde \Gamma$ with $Cl \,  \Gamma \, = \, Cl \, \tilde
\Gamma$
and therefore, each finite topological space can be vizualized
by a minimal graph. For card $X=2$ this minimal graph is
unique.

In  general,  however,   there exists more than one minimal
graph for  one  topology.   For  card  $X  \ge  3$,
 $\{(x_1,x_2),  \, (x_2,x_3),  \,  (x_3,x_1)\}$
and \\* $\{(x_1,x_3),  \, (x_3,x_2), \, (x_2,x_1)\}$
 represent  two different minimal graphs  with  same
transitive closure.  It holds:  The minimal graph representing
 a finite  topological  space  $X$ is unique iff all  the
connected components  of  $X$ possessing more than two  points
are  $T_0$-spaces.\\
\nl
\nl
{\Large
{\bf ACKNOWLEDGEMENTS}}\\
\nl
We thank the referees for valuable comments.
Financial supports from Deutsche Forschungsgemeinschaft and
from the Wis\-sen\-schaftler--Integrations--Programm are
gratefully acknowledged.\\
\nl
\nl
{\Large
{\bf REFERENCES}}\\
\nl
{\small
\* 1. Krasi\'nski, A. (1993). {\it Physics in an inhomogeneous
universe}
\par (ZGUW Warsaw).\\
\* 2. Krasi\'nski, A. (1994). {\it Bibliography on
inhomogeneous
models} Preprint No 273,
\par submitted to Acta Cosmologica.\\
\* 3. Collins, C. B. (1979). {\it Gen. Rel. Grav.} {\bf 10}, 925.\\
\* 4. Krasi\'nski, A. (1983). {\it Gen. Rel. Grav.}
{\bf 15}, 673.
\par Krasi\'nski, A. (1981). {\it Gen. Rel. Grav.} {\bf 13},
1021.\\
\* 5. Szekeres, P. (1975). {\it Commun. Math. Phys.} {\bf 41},
55.\\
\* 6. Schmidt, H.-J. (1982). {\it Astron. Nachr.} {\bf 303},
283.\\
\* 7. Schmidt, H.-J. (1982). {\it Astron. Nachr.} {\bf 303},
227.\\
\* 8. Wolf, J. (1967). {\it Spaces of constant curvature}
(McGraw New York).\\
\* 9. Karlhede, A., and MacCallum, M. A. H. (1982). {\it Gen.
Rel. Grav.}
{\bf 14}, 673.\\
10. Szafron, D. (1981). {\it J. Math. Phys.} {\bf 22}, 543.\\
11. Bona, C., and Coll, B. (1992). {\it J. Math. Phys.} {\bf
33}, 267.\\
12. Paiva, F., Rebouas, M., and MacCallum, M. A. H. (1993).
{\it Class. Quant. \par Grav.} {\bf 10}, 1165.\\
13. Schmidt, H.-J. (1988). {\it J. Math. Phys.} {\bf 29},
1264.\\
14. Schmidt, H.-J. (1987). {\it J. Math. Phys.} {\bf 28},
1928.\\
15. Rainer, M. (1994). {\it The topology of the space of real
Lie algebras
up to \par dimension 4 with applications to homogeneous
cosmological models}
\par (Ph. D. thesis, Universit\"at Potsdam).\\
16. Segal, I. (1951). {\it Duke Math. J.} {\bf 18}, 221.\\
17. Conatser, C. (1972). {\it J. Math. Phys.} {\bf 13}, 196.\\
18. Levy--Nahas, M. (1967). {\it J. Math. Phys.} {\bf 8},
 1211.\\
19. Saletan, E. (1961). {\it J. Math. Phys.} {\bf 2}, 1.\\
20. Schempp, W. (1986). {\it Harmonic analysis on the
Heisenberg nilpotent Lie group} \par (J. Wiley New York).\\
21. Kaplan, A. (1983). {\it Bull. Lond. Math. Soc.} {\bf 15},
35.\\
22. Bianchi, L. (1897). {\it Mem. della Soc. Italiana delle
Scienze  Ser. 3a} {\bf 11}, 267.\\
23. Bianchi, Luigi (1918). {\it Lezioni Sulla Teoria Dei
Gruppi Continui Finiti \par Di Trasformazioni} (Pisa).\\
24. Lie, S., and Engel, F. (1888). {\it Theorie der
Transformationsgruppen}
(Teubner \par Leipzig).
(Remark: Part 3 of this textbook, which contains the
classification
\par
of the $3$-dimensional Lie algebras, appeared in 1893. The 3
volumes
\par
together have more than 2000 pages. Sometimes, this textbook
is cited
\par
without the second author.)\\
25. Lie, Sophus (1891) {\it Differentialgleichungen} (Chelsea
Leipzig).\\
26. Lee, H. C. (1947). {\it J. Math. Pures et Appl.} {\bf 26},
251.\\
27. Ellis, G. F. R., and MacCallum, M. A. H. (1969).
{\it Commun. Math. Phys.} \par {\bf 12}, 108.\\
28. Estabrook, F., Wahlquist, H., and Behr, C. (1968).
{\it J. Math. Phys.} {\bf 9}, 497.\\
29. Davies, P., and Twamley, J. (1993).
{\it Class. Quant. Grav.} {\bf 10}, 931.\\
30. Ambrose, W., and Singer, I. (1958). {\it Duke Math. J.}
{\bf 25}, 647.\\
31. Tricerri, F., and Vanhecke, L. (1983)
{\it Homogeneous structures on Riemannian
\par
manifolds} (Cambridge University Press).\\
32. Christodoulakis, T., and Korfiatis, E. (1992).
{\it J. Math. Phys.} {\bf 33}, 2868.\\
33. Vranceanu, G. (1964).
{\it Le\c cons de G\'eom\'etrie  diff\'erentielle} (Paris).\\
34. Milnor, J. (1976). {\it Adv. Math.} {\bf 21}, 293.\\
35. Rei\ss , A. (1993). {\it Kr\"ummungsinvarianten in
$3$-dimensionalen
homogenen
\par
Riemannschen R\"aumen}  (Staatsexamensarbeit, Universit\"at
Potsdam).\\
36. Slansky, R. (1981). {\it Phys. Reps.} {\bf 79}, 1.\\
37. Ziller, W. (1982). {\it Math. Ann.} {\bf 259}, 351.\\
38. Koike, T., Tanimoto, M., and Hosoya, A. (1994).
{\it J. Math. Phys.} {\bf 35}, 4855.\\
39. Bona, C., and Coll, B. (1994). {\it J. Math. Phys.} {\bf
35}, 873.\\
40. Schmidt, H.-J. (1994).
{\it Why do all the curvature invariants of a gravitational
\par
wave vanish ?} (Preprint Uni Potsdam Math 94/03) to appear: G.
Sarda-
\par nashvili, R. Santilli  (Eds.)
{\it New frontiers in gravitation} (Hadronic Press).\\
41. Nomizu, K. (1979). {\it Osaka J. Math.} {\bf 16}, 143.\\
42. Alias, L., Romero, A., and S\'anchez, M. (1995).
{\it Gen. Relat. Grav.} {\bf 27}, 71.\\
43. Lachieze-Rey, M., and Luminet, J.-P. (1995).
{\it Physics Reports} {\bf 254}, 135-214.}

\end{document}